# Uncertainty in wave hindcasts in the North Atlantic Ocean

Tsubasa Kodaira[1*], Kaushik Sasmal[1**], Rei Miratsu[2], Tsutomu Fukui[2], Tingyao Zhu[2], Takuji Waseda[1]

*1 The University of Tokyo, Graduate School of Frontier Sciences, Kashiwa, Chiba, Japan*
*2 Nippon Kaiji Kyokai (ClassNK), Research Institute, Chiyoda-ku, Tokyo, Japan*

\* *Corresponding author, E-mail addresses:kodaira@edu.k.u-tokyo.ac.jp (T. Kodaira)*
\*\* *Present Affiliation: Technology Centre for Offshore and Marine, Singapore*

**Abstract**

Accurate ocean surface wave knowledge is crucial for ship design. With the significant advancements of model physics and numerical resources, the recent numerical wave hindcast data has a potential to provide environmental conditions for wave load estimation in the ship design process. This study aims to quantify model uncertainty in the state of art numerical wave hindcast products to get insight into the application of wave hindcast for ship design. The model uncertainty is deduced based on the comparison with the wave buoys in the North Atlantic as well as the inter-model comparison with four wave model products. The multiple wave buoys distributed over the Northwest Atlantic and Northeast Atlantic are considered as wave buoy arrays and used for model evaluations. All the four models in general showed very good and similar accuracy in the estimation of the significant wave height $H_s$. The model $H_s$, however, deviates from buoys and also deviate among each other for the extreme wave conditions of $H_s > 10$. It is also found that there is disagreement for the mean wave period $T_{m02}$ between numerical wave models and the wave buoys. More specifically, the models underestimate $T_{m02}$ for extreme wave conditions of $H_s > 10$ in the Northeast Atlantic. The joint probability density function (JPD) of the significant wave height $H_s$ and mean wave period $T_{m02}$ for the extreme wave conditions are also derived from the buoy observations and the hindcast products. Among the JPDs, the total number of samples that satisfies the extreme wave conditions of $H_s > 10$ and $8 < T_{m02} < 14$ varied by 50% although the mean value of the four model products is close to the observation. This study indicates further model improvement is necessary to more accurately reproduce the extreme conditions. For the model improvement as well as the model evaluation, more measurements of the high sea states and their wave spectra information are warranted.

## 1   Introduction

Ocean surface waves are one of the most important environmental conditions for ship design [1]. For the secure ship design, International Association of Classification Societies (IACS) recommend to estimate extreme wave conditions for a design lifetime assuming to be of 25 years, corresponding to the probability of 10[-8] [2],[3]. For this purpose, Standard Wave Data is provided by IACS to ship building community as wave information source to estimate the extreme wave loads. Standard Wave Data are constructed by the wave measurements in the North Atlantic Ocean because the wave climate there is most severe for ships in the global ocean [4]. The joint probability of significant wave height $H_s$ and average zero up-crossing wave period $T_z$ are provided based on the historical wave observations by Voluntary Observing Ships' (VOS), which is compiled as Global Wave Statistics (GWS) and issued by the British Maritime Technology [4]. Although Global Wave Statistics are created based on visual observations accumulated in 130 years, the dataset does not necessarily include the stormy conditions because ship usually avoid encountering the high sea states. Though the previous study [6] extended GWS statistically to include the extreme wave conditions, actual measurements of such stormy condition is preferred because some uncertainty is associated with the statistical extension to estimate the probability of the extreme wave conditions.

After the Global Wave Statistics was published the technological development of ocean wave measurement was significant. Historically, moored wave buoys and wave gauges at the ocean platforms were the standard way to measure the wave conditions with sensors. Routine measurements of ocean surface waves from satellites started with GEOSAT in 1985 and continued the global wave observations without significant interruption [7]. Though the satellite measurements cover the global ocean, measurement is limited to certain location at certain time. Therefore, the satellite measurement does not necessarily include the extreme wave condition occurred. Recently, a new innovative wave observation technology of the extensive distributed wave buoy network is emerging. Massive amount of directional wave buoys has been already distributed over the Pacific Ocean, the Indian Ocean, and also the Atlantic Ocean It was shown that the distributed buoys provide effective data for wave prediction through the data assimilation [8]. The distributed wave buoy network is promising to be information source of new standard wave data for ship design in future. It requires however some time to accumulate the data to cover sufficiently wide range of the wave conditions because the atmosphere and ocean have wide range of temporal variations including the decadal variations.

Another way to obtain vast amount of wave information is to make use of the numerical ocean wave models. Like the weather forecast, sea state has been nowadays routinely forecasted by so called third-generation spectral wave models. The numerical wave models can be also used for the hindcast simulation to obtain comprehensive dataset of ocean waves in target region as well as the entire globe. As far as the model grids can resolve the ocean waves under the stormy weather condition, the high sea state can be estimated though the model accuracy need to be carefully examined. There are several wave hindcast products renowned in the research community. ERA5 is a global reanalysis wave product based on the coupled wave-atmosphere model and data assimilation using ECWAM [9]. IOWAGA is a global hindcast wave field based on reanalysis wind fields and uses WAVEWATCH III® [10]. The data period of ERA5 and IOWAGA is from 1979 onward and 1990 onward, respectively. Since both ERA5 and IOWAGA are the global wave hindcast products and the data period is more than two decades, it is perhaps practical to make use of these hindcast products to redefine the design wave. Unlike the limited in-situ observations, the hindcast does not miss the stormy conditions though the peak is sometimes not well reproduced as highlighted by the previous study [11]. Further consideration on the possible update of the design wave information requires comprehensive assessment of wave hindcast products.

The evaluation of the wave model products can be often found in the literature. Both in situ and satellite measurements were used as reference to provide the error metrics of the wave model results. The metrics used are such as bias, correlation, and RMSE, and sometimes Taylor diagram is used as succinct visualization tool for these error metrics [12]. Regarding the significant wave height, agreement between the model estimates and buoy measurements are high such that correlation coefficient often exceed 0.9. The leading factor of the model accuracy was also elucidated as the accuracy of the wind forcing for the wave model [13]. There are much less but several literatures on the model accuracy of the mean wave period compared to the significant wave height [14].

The error metrics mentioned above can be used to discuss the wave climates but are not necessarily effective to characterize the model ability to estimate the extreme conditions. For example, the previous study showed performance of three wave hindcasts in the North Atlantic Ocean is very similar to each other for non-extreme conditions, but the difference increases significantly such that the difference in significant wave height reached to 2.5 meters [15]. More recently, the global wave model was run with 12 different wind forcing, and the results showed each of the wave hindcasts reproduced the average sea states similarly, but large discrepancies are found for very high sea state conditions [16]. How well the extreme conditions estimated in the wave hindcast products are important to make use of the dataset for ship design in the context of the extreme value analysis such as deriving environmental contours used in structural reliability analysis [17].

This study aims to quantify the uncertainty in wave hindcast products based on the comparison of the multiple wave hindcast products for the North Atlantic. The validation of the hindcast products is also conducted by using the wave buoy observations distributed over the Northwest Atlantic and Northeast Atlantic. Bulk wave statistics of significant wave height and mean wave period are used for the comparison. Since the extreme wave conditions are more important to estimate the extreme wave loads on the ship, the extreme conditions from the multiple wave hindcasts are compared and evaluated in detail. This study examines the extreme wave conditions regardless of the ship locations. For the extreme wave load estimation in practice, ship avoidance of stormy conditions needs to be considered additionally. This topic is discussed in detail in the companion paper of Miratsu et al.[18].

## 2 Data

### 2.1 Wave hindcast data

We analyzed 25 years of wave hindcast data and wave reanalysis data for GWS 8, 9, 15 and 16 areas in the North Atlantic (Figure 1). The description for each of four products are provided below. Numerical model setup of the multiple wave hindcast and reanalysis data are summarized in Table.1.

The in-house TodaiWW3-NK is an Atlantic Ocean regional hindcast wave model using WAVEWATCH III® and NCEP-CFSR as wind forcing. The spatial resolution is 0.20 x 0.25 (Lat x Lon), and the spectral resolutions are 35 frequency bins and 36 directional bins. The frequency range resolved in the model is from 0.04 to 1.05 Hz. For model physics, the ST4 package of Ardhuin et al. [6] was used with $β_{max}$ = 1.33. The model was forced by NCEP/CFSR hourly wind and was integrated for 25 years in hindcast mode (1994-2018). The wave hindcast data are distributed with the interval of one hour. Further information of the numerical model setup can be found in Sasmal et al. [19].

ERA5 is a global reanalysis wave product based on the coupled wave-atmosphere model and data assimilation based on the ocean wave model ECWAM. The spatial resolution is 0.36 x 0.36 (Lat x Lon), and the spectral resolutions are 24 directional bins and 30 frequency bins. The spatial resolution of the distributed dataset is lowered to 0.5 x 0.5 (Lat x

Lon) and so the data used in this study. The frequency range resolved in the model is from 0.0345 to 0.55 Hz. Satellite altimeter data are assimilated in ERA5 for correction of Hs. The wave hindcast data are distributed with the interval of one hour. Further information of the numerical model setup can be found in Hersbach et al. [9].

IOWAGA is a global hindcast wave field based on reanalysis wind fields and uses WAVEWATCH III®. For IOWAGA, the WAVEWATCH III model was forced by either CFSR or ERA5 winds. Each of the two products are named in this study as IOWAGA(CFSR) and IOWAGA(ERA5). The spatial resolution is 0.5 x 0.5 (Lat x Lon), and the spectral resolutions are 24 directional bins and 30 frequency bins. The frequency range resolved in the model is from 0.034 to 0.95 Hz. The wave hindcast data are distributed with the interval of three hours. Further information can be found in Rascle et al. [20].

**Table 1 Wave hindcast data**

| Name | TodaiWW3-NK | ERA5 | IOWAGA(CFSR) | IOWAGA(ERA5) |
|---|---|---|---|---|
| Institution | The University of Tokyo | ECMWF | IFREMER | IFREMER |
| Model Domain | Atlantic | Global | Global | Global |
| Spatial Res. | 0.2°x 0.25° | 0.5°x 0.5° | 0.5°x 0.5° | 0.5°x 0.5° |
| Spectral Res. | 36 directional bins 35 frequency bins | 24 directional bins 30 frequency bins | 24 directional bins 31 frequency bins | 24 directional bins 31 frequency bins |
| Wind Forcing | CFSR | Coupled | CFSR | ERA5 |

## 2.2 Wave observation data

Wave buoy data are obtained from the Copernicus Marine Environment Monitoring Service (CMEMS) in situ Thematic Assembly Centre (TAC). Fig. 1 provides a geographical map of the wave buoys used in this study. The buoys are owned by one of the following institutions: NDBC/NOAA US, WHOI US, DFO Canada, Marine Institute Ireland, and UK Metoffice. The identifier for each buoy from World Meteorological Organization (WMO) is listed as legend in Fig.1. The time coverage of each of the wave buoys can be found in Fig. 2. Some of the buoys measured most of the data period of 25 years from 1994 to 2018 examined in this study.

The wave buoy locations are distributed to form two clusters in the western and eastern part of the study region. We therefore regard the two clusters as two sets of wave buoy array and compare those data with the wave hindcast products. The clustering of the buoys should provide higher probability to capture the high sea states such as the one caused by intense storm passage. The data from each buoy was merged after avoiding the noisy data based on the quality control flag.

All the wave buoys provide the significant wave height data, but mean wave period is not provided from all the buoys (Fig.2). Moreover, although the mean wave period is provided from all the buoys located over the Northeast Atlantic, but the period is provided with the resolution of one second. Sine the buoys from NDBC/NOAA provide frequency wave spectra information, the data are used to discuss how to estimate the mean wave period in detail.

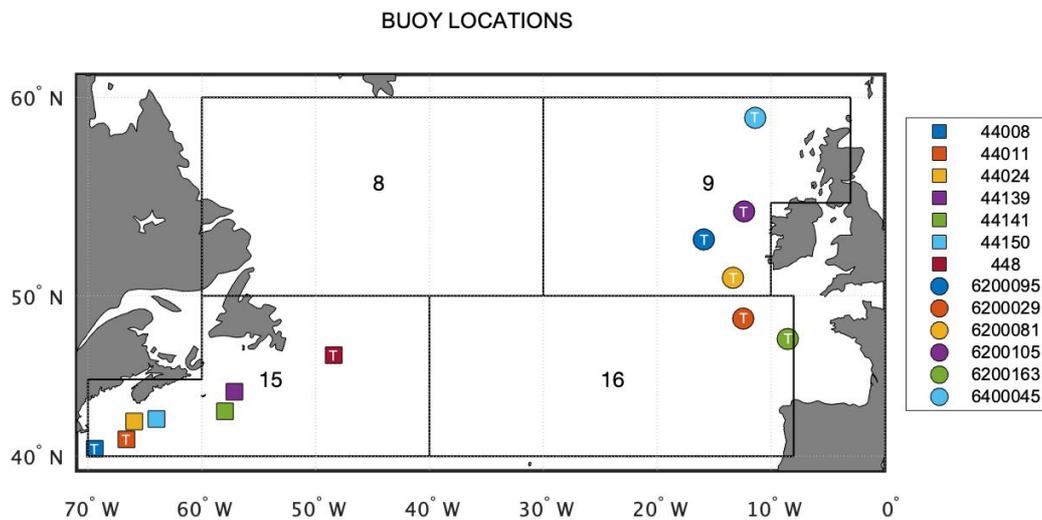

Fig. 1 Moored wave buoy locations over the GWS area 8, 9, 15, 16 defined in *IACS Rec. No. 34* in the North Atlantic. The legend shows WMO identifier. The upper case T in some of the marker indicates the buoys that provide mean wave periods.

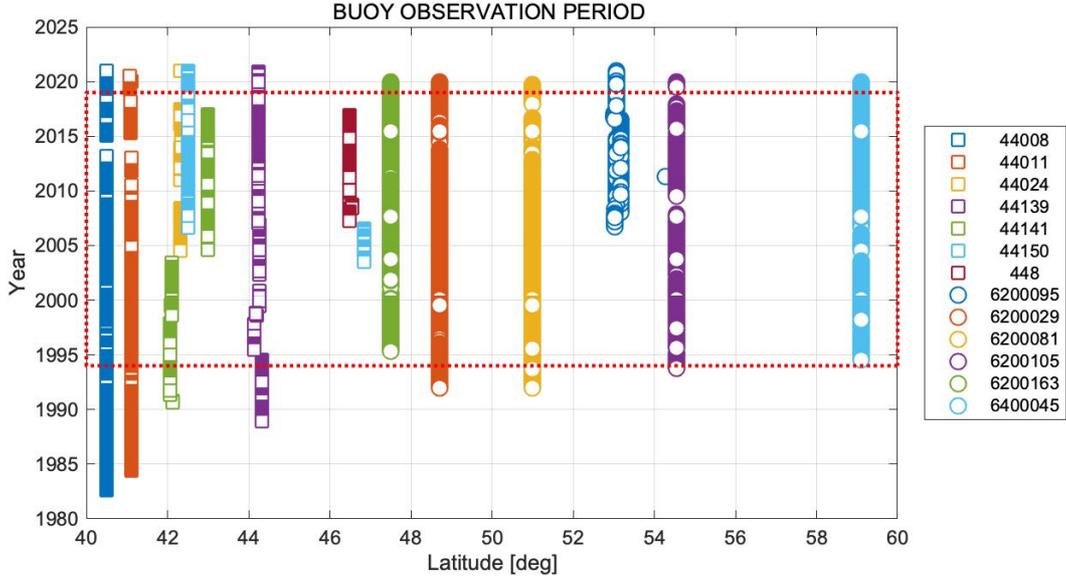

Fig. 2 Period of the wave measuremet by the wave buoys shown in Fig.1. The legend shows WMO identifier.

## 3   Results

### 3.1   Significant Wave Height

The significant wave height measured in the Northwest Atlantic is first examined. Based on the bivariate histogram for comparison between wave buoy and each model (Fig. 3), all the four model products are in good agreement with the observational data. The correlation coefficients are close to 0.95 although the models slightly underestimate the in-situ observations. Some differences are however found for the relatively high sea state conditions ($H_s > 10$). TodaiWW3-NK and IOWAGA (CFSR) successfully reproduces the largest significant wave height around 15m. ERA5 does not reproduce these conditions and results in underestimation. IOWAGA (ERA5) reproduced those high sea states, but often overestimate the wave height such as the result of 16m wave height, while the wave buoy measured only 10m.

Over the Northeast Atlantic, sea states are more severe than the Northwest Atlantic. The maximum significant wave height measured in the Northeast Atlantic is 19m while it was less than 16m in the Northwest Atlantic. Based on the bivariate histogram (Fig. 4), TodaiWW3-NK and IOWAGA (CFSR) shows excellent agreement for the wide range of $H_s$. The linear regression lines are almost the same as one-to-one lines, which indicates the models perform somehow better in the Northeast Atlantic than Northwest Atlantic. The correlation coefficients exceed 0.95 for all the four wave hindcast products. ERA5 however consistently underestimate $H_s$. IOWAGA (ERA5) on the other hand again tends to overestimate the high sea states for the Northeast Atlantic.

To further rank the hindcast products, typical error metrics are compared to each other using so called Taylor diagram. This way of visualization is proposed as a graphical approach to describe how well one dataset matches the other in terms of the error metrics: standard deviation, correlation coefficient, and the centered root-mean square difference (RMSD) [12]. Taylor diagrams for the significant wave height measured in the Northwest Atlantic and Northeast Atlantic are shown in Fig. 5 as well as the numeric information listed in Table.2. As a result, ERA5 shows slightly better performance than the other products as far as the correlation coefficient and the centered RMSD are used. ERA5 has however relatively large negative bias especially for the Northeastern Atlantic case as described in the previous paragraphs (Table 2).

To further examine the model performance including the high sea state conditions, Quantile-Quantile plot (QQ-plot) is used (Fig. 6). For the Northwest Atlantic, very good agreement between the buoys and models is found up to 11m for all the products. Beyond 11m, IOWAGA(ERA5) starts to overestimate deviate to the upper while ERA5 deviates to the lower. TodaiWW3-NK and IOWAGA(CFSR) showed better and similar model skills even for the high sea state conditions. It is also clearly shown in the QQ-plot that maximum difference among the four wave hindcast products increase with Hs. The difference reaches almost 4m.

For the Northeast Atlantic, ERA5 start to deviate to the lower from the one-to-one line in QQ-plot and from the other three hindcast products after 8m (Fig. 6). IOWAGA(ERA5) next deviates to the upper after 12m. TodaiWW3-NK and IOWAGA(CFSR) finally deviates from the one-to-one line after 15m. IOWAGA(CFSR) perform slightly better than TodaiWW3-NK, which indicates the probability function of $H_s$ measured by the wave buoys in the Northeast

Atlantic was reproduced most well by IOWAGA(CFSR). The maximum difference among the four wave hindcast products also increases with $H_s$ in the Northeast Atlantic and reaches 5m.

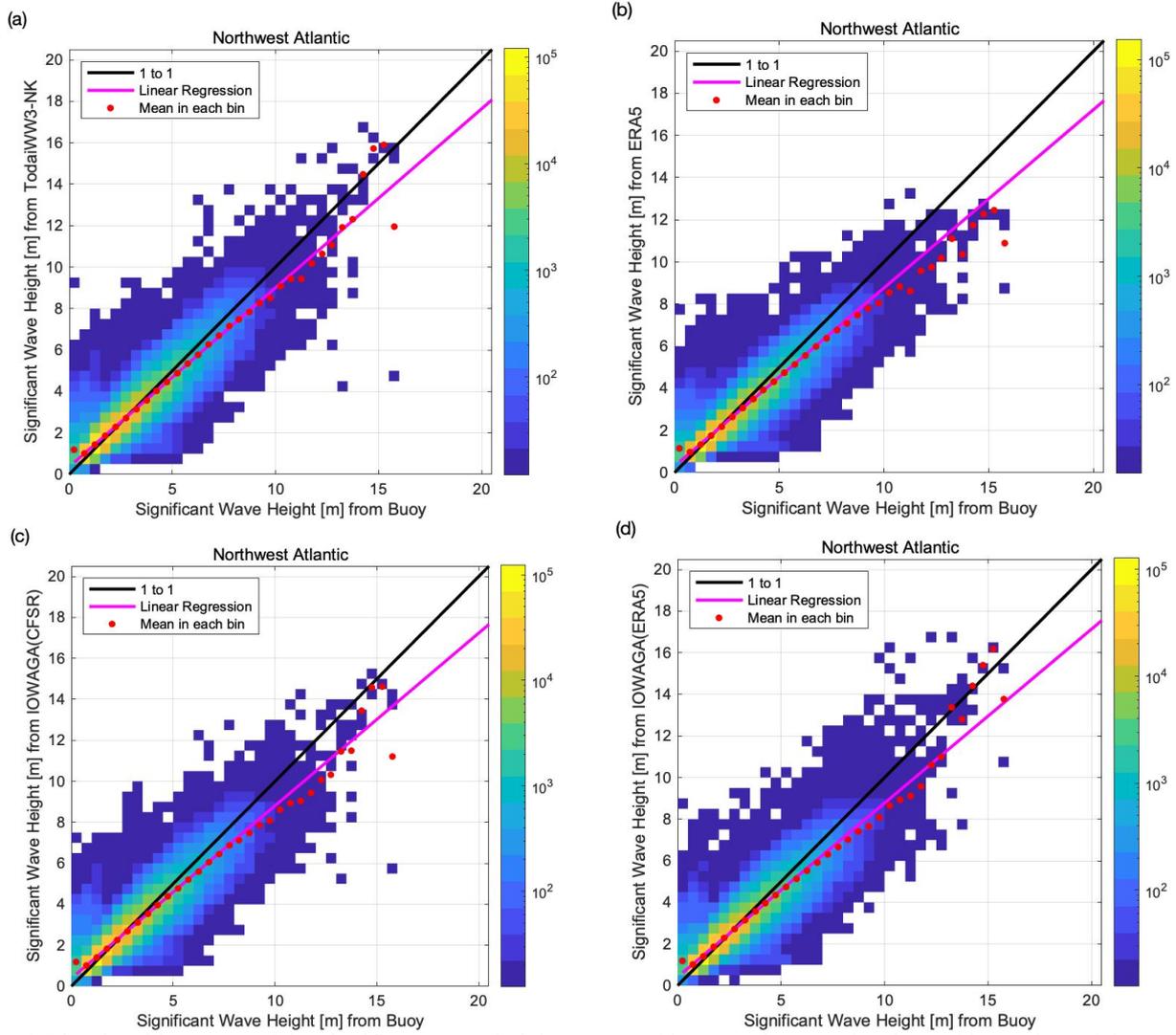

**Fig. 3** Bivariate histogram of the significant wave height measured by wave buoy and estimated by each model products for the Northwest Atlantic.

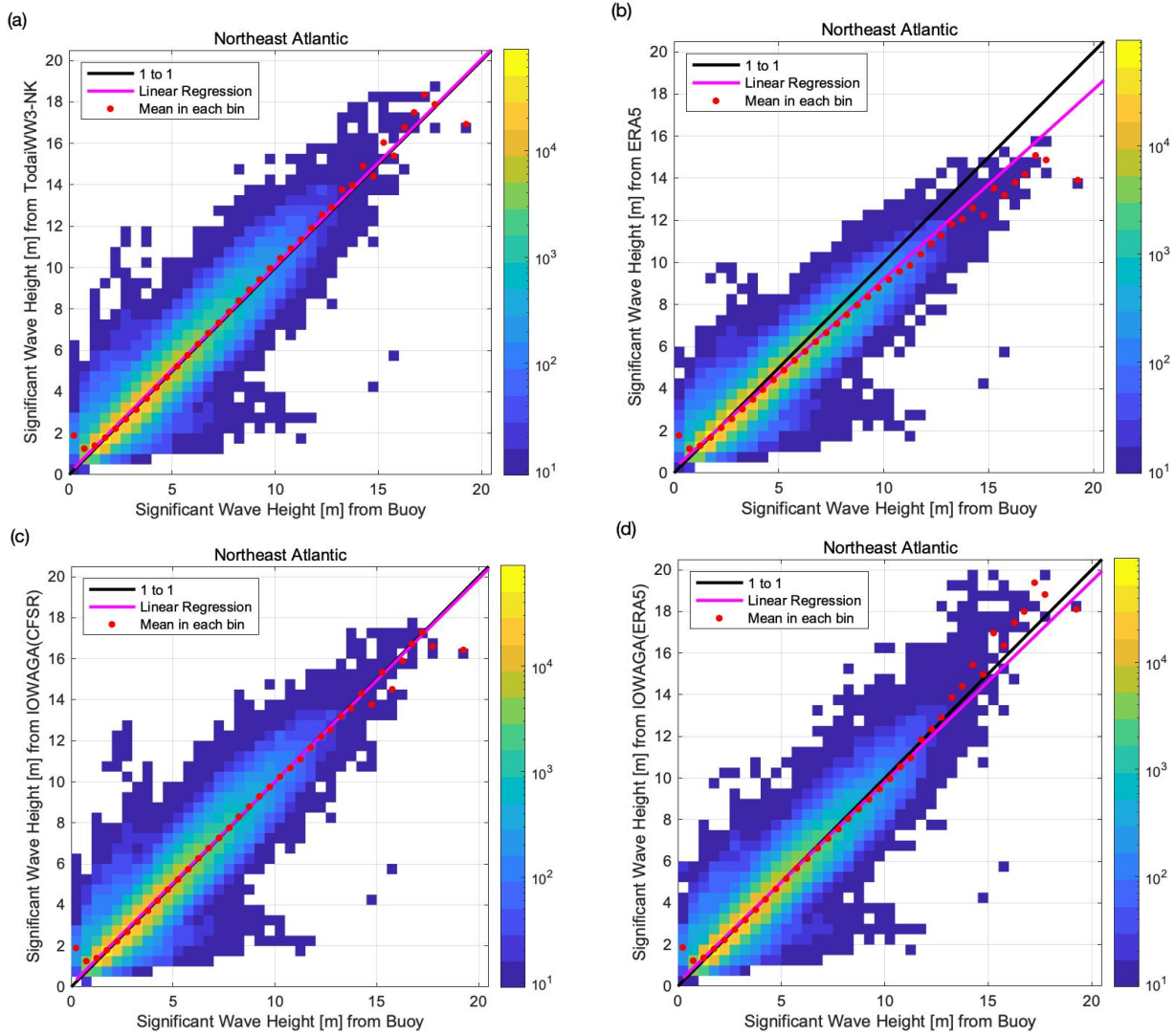

**Fig. 4** Bivariate histogram of the significant wave height measured by wave buoy and estimated by each model products for the Northeast Atlantic.

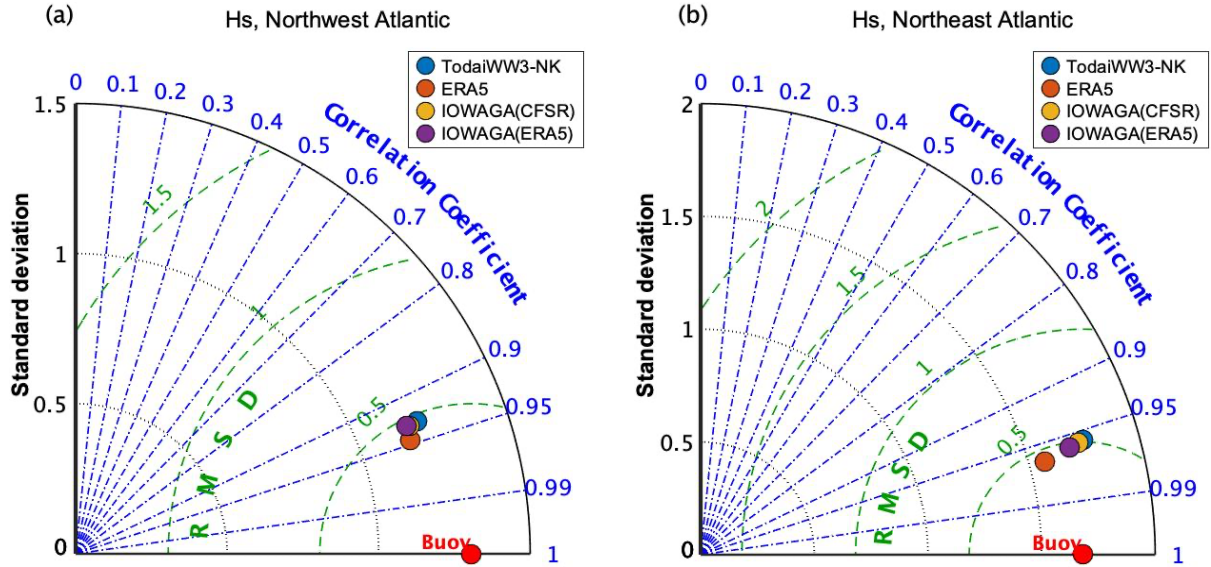

Fig. 5 Taylor diagram of the significant wave heigt to evaluate the four wave hindcast products based on the observation by wave buoy array. The left panel and right panel is for the Northwest Atlantic and Northeast Atlantic, respectively.

Table 2  Model evaluation metrics for the significant wave height.

|  |  | Buoy-Array | TodaiWW3-NK | ERA5 | IOWAGA (CFSR) | IOWAGA (ERA5) |
|---|---|---|---|---|---|---|
| Northwest Atlantic | Mean | 2.17 | 2.24 | 2.14 | 2.21 | 2.22 |
|  | Standard deviation | 1.31 | 1.21 | 1.17 | 1.18 | 1.17 |
|  | Root mean square difference | - | 0.48 | 0.43 | 0.47 | 0.48 |
|  | Correlation coefficient | - | 0.931 | 0.946 | 0.932 | 0.932 |
| Northeast Atlantic | Mean | 3.19 | 3.27 | 3.08 | 3.27 | 3.24 |
|  | Standard deviation | 1.68 | 1.76 | 1.57 | 1.74 | 1.69 |
|  | Root mean square difference | - | 0.51 | 0.45 | 0.50 | 0.48 |
|  | Correlation coefficient | - | 0.957 | 0.965 | 0.958 | 0.959 |

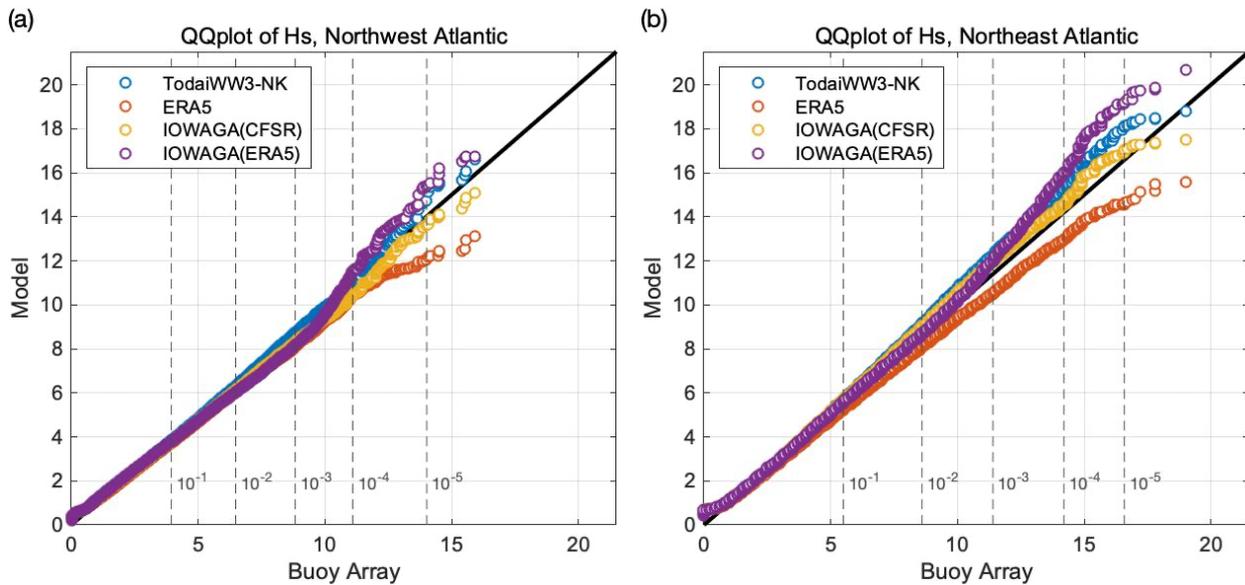

Fig. 6  Quantile-quantile plot of the significant wave heigt to evaluate the four wave hindcast products based on the observation by wave buoy array. The left panel and right panel is for the Northwest Atlantic and Northeast Atlantic, respectively.

## 3.2 Mean Wave Period

We next compare the model results of mean wave period with the buoy observations. This comparison of the mean wave period is not as straightforward as the significant wave height. To highlight the practical difficulty of the comparison, data from NDBC Buoy #44011 and corresponding predictions by TodaiWW3-NK are examined detailly and carefully. This buoy and TodaiWW3-NK were chosen because wave spectra information is fully available in both the data.

The mean wave period discussed first is spectral moment period $T_{m02}$ and $T_{m01}$. One of the mean wave periods, $T_{m02}$ is defined as

$$T_{m02} = \left(\int_{f_1}^{f_2} S(f)df \Big/ \int_{f_1}^{f_2} f^2 S(f)df\right)^{1/2} \quad (Eq.1)$$

where $S(f)$ is omni-directional spectra, $f$ is frequency in Hz, $f_1$ is lower frequency limit, and $f_2$ is upper frequency limit. Some attention is necessary to the frequency range from $f_1$ to $f_2$ because $T_{m02}$ depends on the range. The frequency range of NDBC Buoy #44011is from 0.03 to 0.4, while the model is from 0.03 to 1.0. Two sets of $T_{m02}$ are thus produced from the model results by changing the upper frequency limit $f_2$ from 1.0 to 0.4. With $f_2 = 1.0$, the mean wave periods from the model are found shorter compared to the buoy observation (Fig. 7a). The agreement improved when the integration range is set to the same with each other by setting $f_2$ to 0.4 (Fig. 7b). More specifically, the mean value of $T_{m02}$ is calculated as 5.98, 5.34, and 5.92 for wave buoy, TodaiWW3-NK($0.03 < f < 1.0$), and TodaiWW3-NK ($0.03 < f < 0.4$), respectively. The improvement is however limited up to intermediate values, and no clear improvement can be found around 10s (Fig. 7). These results indicate that the corresponding wave spectra are not well reproduced by the model.

Similar observation can be made for another mean wave period T₀ₘ₁ defined as,

$$T_{m01} = \int_{f_1}^{f_2} S(f)df \Big/ \int_{f_1}^{f_2} fS(f)df \quad (Eq.1).$$

Two sets of $T_{m01}$ are again produced from the model results by changing the upper frequency limit $f_2$ from 1.0 to 0.4. With $f_2 = 1.0$, the mean wave periods from the model are also shorter compared to the buoy observation (Fig. 8a), but the degree of underestimation is smaller than the case of $T_{m02}$. More specifically, mean values of $T_{m01}$ are 6.40 and 5.98, for wave buoy and TodaiWW3-NK($0.03 < f < 1.0$), respectively. When the $f_2$ is set to 0.4, the mean value of $T_{m01}$ is calculated as 6.33, which is closer to the buoy observation. The improvement is again limited up to the intermediate values, and no clear improvement can be found around 10s (Fig. 8).

The results above indicate that when the mean wave periods are relatively large the change in upper frequency limit $f_2$ from 1.0 to 0.4 does not influence much on the calculation of $T_{m01}$ and $T_{m02}$. The selection of the upper frequency limit $f_2$ is further considered by examining how the sensitivity changes with the significant wave height. The relation between the significant wave height and ratio of two mean wave periods $T_{m02}'/T_{m02}$ and $T_{m01}'/T_{m01}$ are thus examined. The dash denotes the values calculated with $f_2 = 0.4$, while the other one is calculated with $f_2 = 1.0$. The relation between $T_{m02}'/T_{m02}$ and significant wave height $H_s$ from TodaiWW3-NK clearly shows that change in $f_2$ becomes significantly smaller with $H_s$. If the focus is limited to the samples with $H_s > 5$, the value of $T_{m02}'/T_{m02}$ is almost always less than 1.05. Similar observation can be made for $T_{m01}'/T_{m01}$(Fig. 9).

For the reference, the effect of the truncation of integration range by specifying $f_2$ is calculated based on the Pierson Moskowitz (PM) spectrum. This time, regarding $T_{m02}'/T_{m02}$ or $T_{m01}'/T_{m01}$, the tilde is for the values calculated with finite integration range, while the other one is calculated based on the analytical formula of PM spectra ($T_z = T_{m02} = 3.55 \sqrt{H_s}$, $T_{m01} = 1.087 T_{m02}$) [21]. It is shown that the error in the calculation of mean wave period $T_{m02}$ and $T_{m01}$ are negligible with $f_2 = 1.0$, but noticeable with $f_2 = 0.4$. It is also found that $T_{m02}'/T_{m02}$ and $T_{m01}'/T_{m01}$ based on the model results are often slightly larger than the analytical results based on PM spectra. The influence of the truncation can be illustrated by some power spectra from the model data. In Fig.10, subset of power spectra corresponding to the case of $H_s = 3$ and $T_{m02}'/T_{m02} = 1.1$ are shown. These examples clearly show that the double peak nature of the power spectra results in the larger value of $T_{m02}'/T_{m02}$.

We finally conduct the comparison of the mean wave period from the model and wave buoy array. Only the high sea state conditions are examined ($H_s > 10$) because these conditions are important for the ship design and because the uncertainty of the integration range for mean wave periods is smaller for high sea states as we examined above. With this condition of $H_s > 10$, only the wave buoy array in the Northeast Atlantic are compared to the model results because the number of samples for the high sea state are very small in the wave buoy array in the Northwest Atlantic. The comparison of the buoy observation in the Northeast Atlantic with each model results indicate that all of four wave hindcast tend to underestimate the mean wave periods (Fig.11).

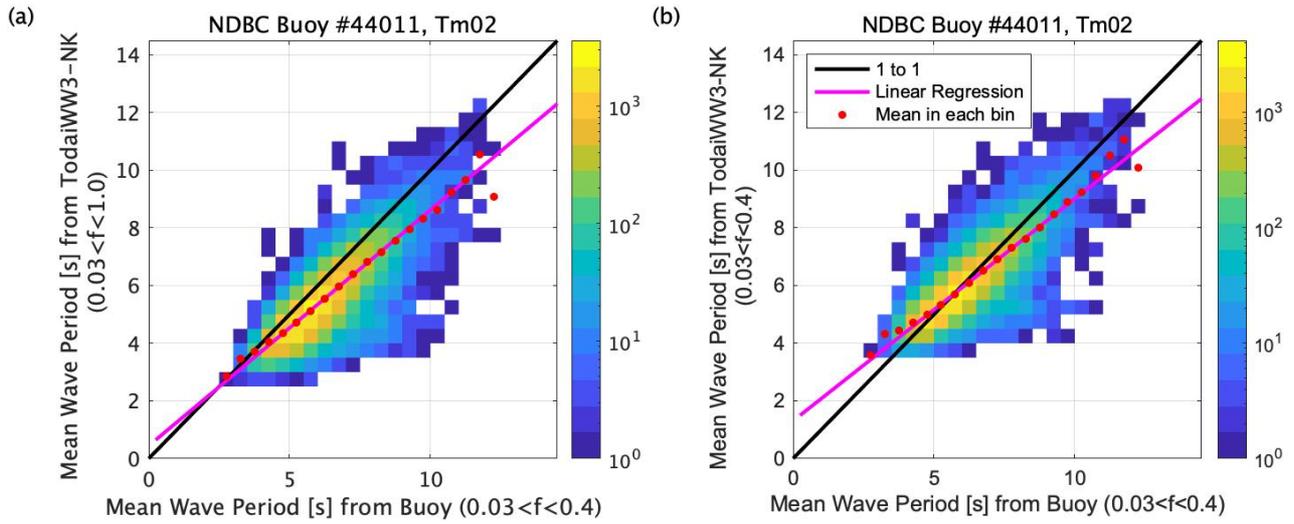

**Fig. 7** Bivariate histogram of the mean wave period $T_{m02}$ measured by wave buoy and estimated by TodaiWW3-NK. The left and right panel shows the results with the frequency range from 0.03 to 1.0 and from 0.03 to 0.4, respectively.

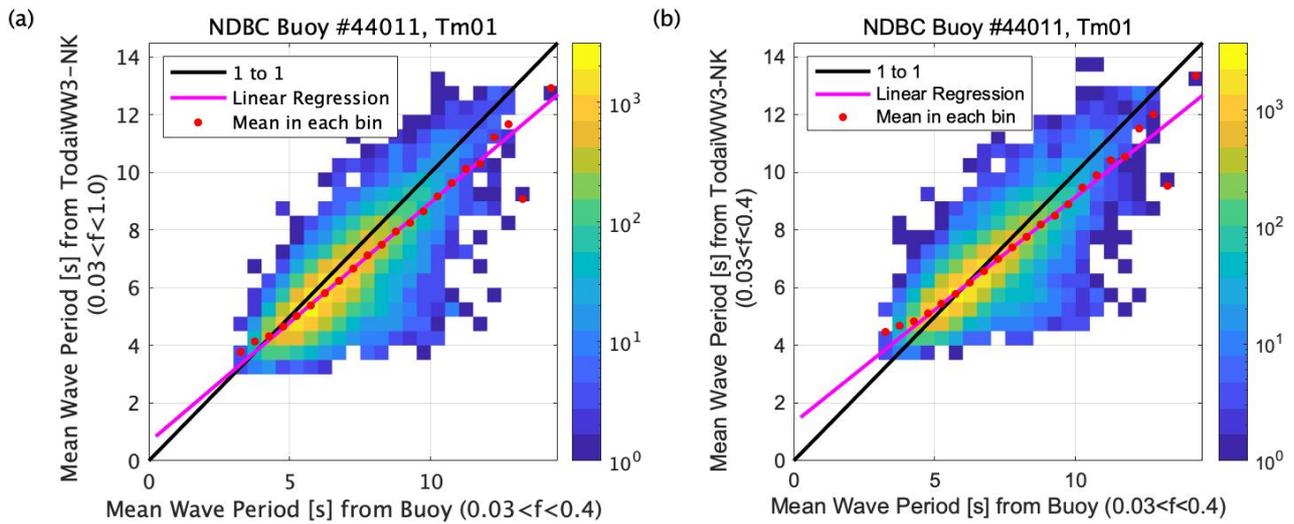

**Fig. 8** Bivariate histogram of the mean wave period $T_{m01}$ measured by wave buoy and estimated by TodaiWW3-NK. The left and right panel shows the results with the frequency range from 0.03 to 1.0 and from 0.03 to 0.4, respectively.

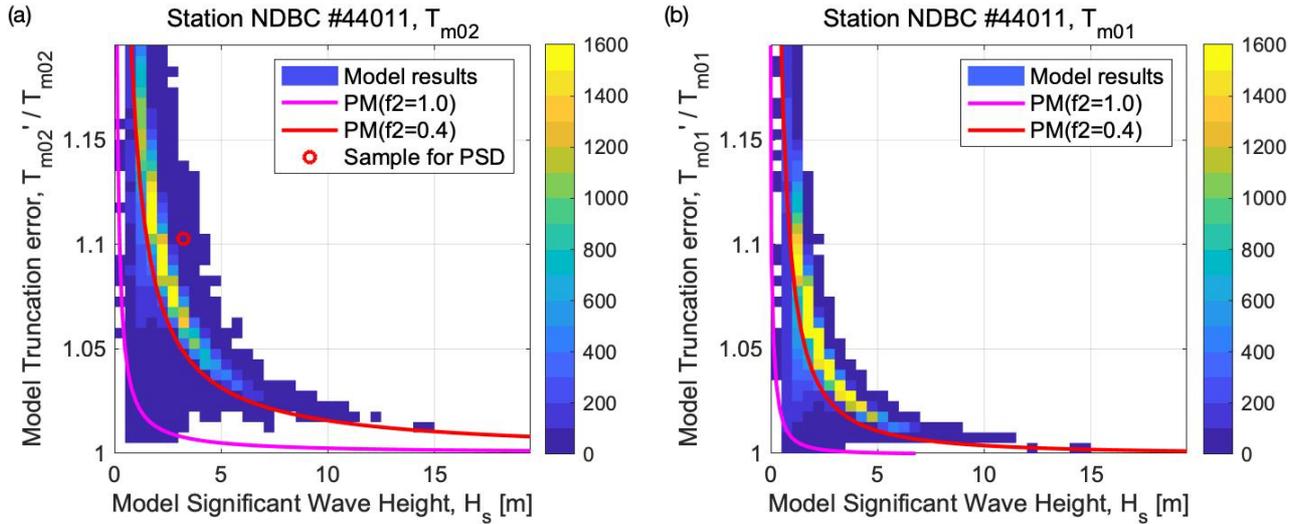

**Fig. 9** Bivariate histogram of the model significant wave height $H_s$ and the truncation error in the mean wave period originated in the integration range of the spectral moment $m_2$. All the data shown here are based on the output from TodaiWW3-NK at the position of the NDBC wave buoy #44011.

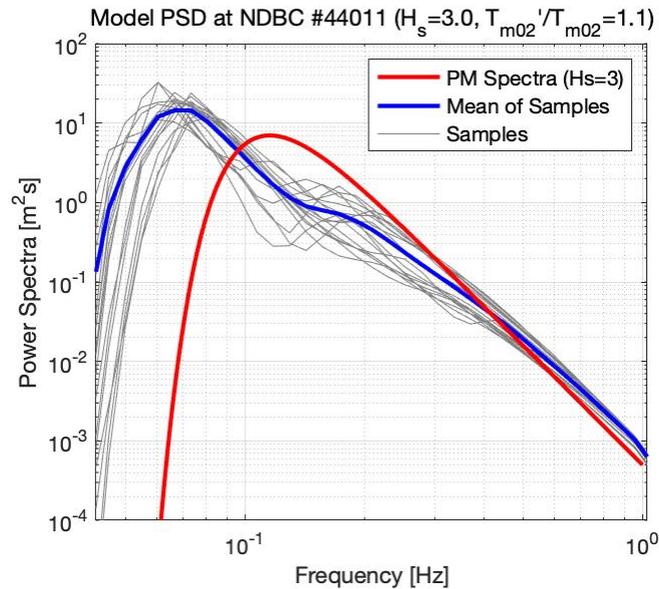

**Fig. 10** Frequency wave spectra from TodaiWW3-NK for the position of the NDBC wave buoy #44011 (black lines). The spectra are chosen by the following conditions: $3 \leq H_s < 3.1$ and $1.1 \leq T_{m02}'/T_{m02} < 1.125$. The blue line indicates the mean of the spectra. The red line shows the Pierson Moskowitz spectrum with the condition of $H_s = 3$.

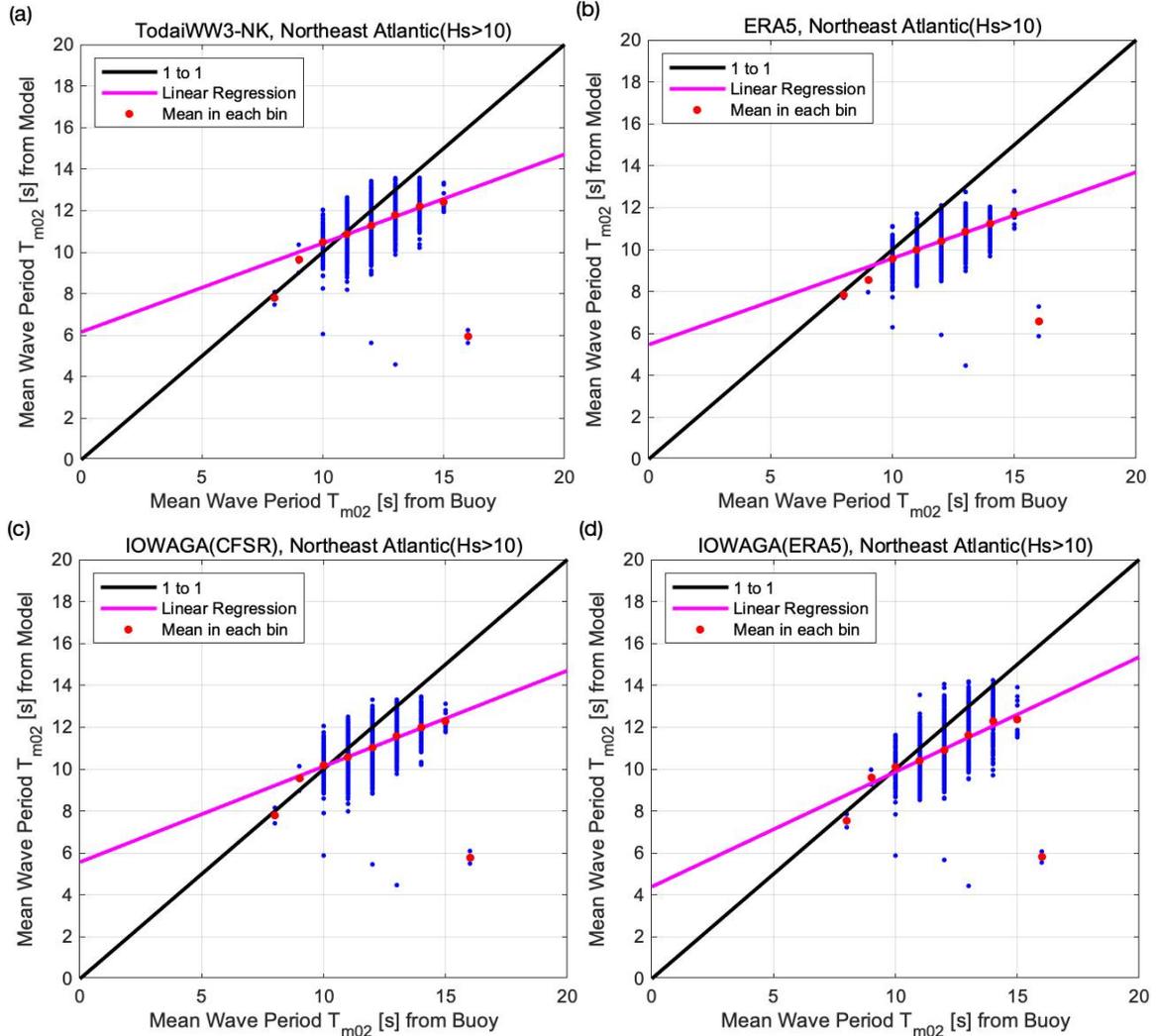

**Fig. 11** Scatter plot of the mean wave periods measured by the wave buoy array in the Northeast Atlantic and corrsponding model results from four wave hind cast products. The blue dots indicate the data when the significant wave height exceeds 10m. The black line is the one-to-one line, while the magenta line indicates the linear least square fit for the data. The red dots show the mean value of the model results in each bin.

### 3.3 Joint Probability of the Significant Wave Height and Mean Wave Period

The wave load generally increases with the significant wave height, but also depends significantly on the wave period. The joint probability function of $H_s$ and $T_{m02}$, often called as Scatter Diagram in the shipbuilding community, is therefore necessary to estimate the wave loads for ship design. In this subsection, part of Scatter Diagram from the multiple wave hindcast are compared to each other. The focus is on the stormy condition that is important for the estimation of the possible maximum wave loads during lifetime of ship. More specifically, we investigate the uncertainty in how models reproduce very high sea states. The companion paper [18] showed ranges of significant wave height and mean wave period that cause the extreme wave loads (see Fig. 8 of [18]). Since the key wave conditions are specified as $10 < H_s < 16$ and $8 < T_{m02} < 14$, this part of Scatter Diagram is discussed. The part of Scatter Diagram is hereafter referred as Scatter Diagram for Extreme conditions (SDE).

The SDE from the wave buoy array and from the multiple wave hindcast products are shown in Table 3 Since buoy array in the Northwest Atlantic collected 100 times less samples than the buoy array in Northeast Atlantic, only the data from the buoy array in Northeast Atlantic are presented. Some important facts can be readily found by comparing the tables. First, the total number of the samples in SDE from the multiple wave hindcast products differ significantly. The maximum number is 7526 for TodaiWW3-NK and the minimum is 2545 for ERA5, while the number of samples from the wave buoys is 4997. These values indicate that the occurrence of the stormy conditions strongly depend on the numerical wave model and its configuration.

Second, all SDEs show the peak value of the mean wave period increases as $H_s$ increases, which is consistent with the Fetch law. It should be noted that the maximum number of samples in each $H_s$ bin is sometimes found at longer mean wave period in the SDE based on the buoy array observation than the model results. Since we confirmed in the previous subsection that the frequency range does not matter in the calculation of $T_{m02}$, the difference in the $T_{m02}$ is not caused by the data processing. Either the model fails to reproduce long waves properly, or the buoy fails to measure the low frequency waves properly. The latter is possible if the buoy uses the accelerometer. The discrepancy is however limited to around 1 s and may not exceed 2 s.

Third, IOWAGA(ERA5) provides intermediate number of samples in total, but the number of samples is more than double of the other data source for extremely stormy conditions with Hs > 15. This is consistent with the QQ plot for $H_s$ (Fig. 6) as the line for IOWAGA(ERA5) transect the lines of TodaiWW3-NK and IOWAGA(CFSR) around $H_s = 10$. This result implies that the model accuracy depends on each storm event and therefore highlights the complexity of the model intercomparison.

**Table 3** The joint probability function of the significant wave height $H_s$ and mean wave period $T_{m02}$ for extreme wave conditions. The tables below are based on the buoy observations and wave hindcast products and the number in each cell indicates the number of samples. The bold face is used for the maximum value in each bin of $H_s$.

| Obs NEA | | Hs | | | | | | | | | | |
|---|---|---|---|---|---|---|---|---|---|---|---|---|
| | | 10 | 11 | 12 | 13 | 14 | 15 | 16 | 17 | 18 | 19 | 20 | Total |
| $T_{m02}$ | 10 | **424** | 39 | 1 | 1 | 0 | 0 | 0 | 0 | 0 | 0 | 0 | 465 |
| | 11 | **1444** | **461** | 78 | 22 | 0 | 0 | 0 | 0 | 0 | 0 | 0 | 2005 |
| | 12 | 768 | 568 | **291** | 108 | 28 | 2 | 3 | 0 | 0 | 0 | 0 | 1768 |
| | 13 | 176 | 148 | 125 | **89** | **57** | **22** | 2 | 1 | 0 | 0 | 0 | 620 |
| | 14 | 21 | 26 | 27 | 19 | 15 | 15 | **10** | **4** | **2** | 0 | 0 | 139 |
| | Total | 2833 | 1242 | 522 | 239 | 100 | 39 | 15 | 5 | 2 | 0 | 0 | 4997 |

| TodaiWW3 NEA | | Hs | | | | | | | | | | |
|---|---|---|---|---|---|---|---|---|---|---|---|---|
| | | 10 | 11 | 12 | 13 | 14 | 15 | 16 | 17 | 18 | 19 | 20 | Total |
| $T_{m02}$ | 10 | **1901** | 238 | 29 | 0 | 0 | 0 | 0 | 0 | 0 | 0 | 0 | 2168 |
| | 11 | 1696 | **1486** | **552** | 101 | 10 | 0 | 0 | 0 | 0 | 0 | 0 | 3845 |
| | 12 | 168 | 267 | 347 | **326** | **176** | 45 | 3 | 0 | 0 | 0 | 0 | 1332 |
| | 13 | 11 | 10 | 8 | 21 | 14 | **53** | **27** | **23** | 7 | 1 | 0 | 175 |
| | 14 | 0 | 0 | 0 | 0 | 0 | 0 | 0 | 3 | **3** | 0 | 0 | 6 |
| | Total | 3776 | 2001 | 936 | 448 | 200 | 98 | 30 | 26 | 10 | 1 | 0 | 7526 |

| ERA5 NEA | | Hs | | | | | | | | | | |
|---|---|---|---|---|---|---|---|---|---|---|---|---|
| | | 10 | 11 | 12 | 13 | 14 | 15 | 16 | 17 | 18 | 19 | 20 | Total |
| $T_{m02}$ | 10 | **908** | 214 | 47 | 12 | 1 | 0 | 0 | 0 | 0 | 0 | 0 | 1182 |
| | 11 | 632 | **300** | **184** | **55** | 9 | 2 | 1 | 0 | 0 | 0 | 0 | 1183 |
| | 12 | 42 | 25 | 42 | 29 | **35** | **6** | **1** | 0 | 0 | 0 | 0 | 180 |
| | 13 | 0 | 0 | 0 | 0 | 0 | 0 | 0 | 0 | 0 | 0 | 0 | 0 |
| | 14 | 0 | 0 | 0 | 0 | 0 | 0 | 0 | 0 | 0 | 0 | 0 | 0 |
| | Total | 1582 | 539 | 273 | 96 | 45 | 8 | 2 | 0 | 0 | 0 | 0 | 2545 |

| IOWAGA-C NEA | | Hs | | | | | | | | | | |
|---|---|---|---|---|---|---|---|---|---|---|---|---|
| | | 10 | 11 | 12 | 13 | 14 | 15 | 16 | 17 | 18 | 19 | 20 | Total |
| $T_{m02}$ | 10 | **1963** | 395 | 40 | 2 | 0 | 0 | 0 | 0 | 0 | 0 | 0 | 2400 |
| | 11 | 1392 | **1132** | **491** | 109 | 10 | 1 | 0 | 0 | 0 | 0 | 0 | 3135 |
| | 12 | 168 | 196 | 278 | **245** | **124** | 27 | 6 | 3 | 0 | 0 | 0 | 1047 |
| | 13 | 4 | 6 | 5 | 15 | 12 | **27** | **21** | **12** | **1** | 0 | 0 | 103 |
| | 14 | 0 | 0 | 0 | 0 | 0 | 0 | 0 | 0 | 0 | 0 | 0 | 0 |
| | Total | 3527 | 1729 | 814 | 371 | 146 | 55 | 27 | 15 | 1 | 0 | 0 | 6685 |

| IOWAGA-E NEA | | Hs | | | | | | | | | | |
|---|---|---|---|---|---|---|---|---|---|---|---|---|
| | | 10 | 11 | 12 | 13 | 14 | 15 | 16 | 17 | 18 | 19 | 20 | Total |
| Tm02 | 10 | 1705 | 300 | 38 | 6 | 3 | 0 | 0 | 0 | 0 | 0 | 0 | 2052 |
| | 11 | 886 | 750 | 373 | 136 | 21 | 4 | 1 | 0 | 0 | 0 | 0 | 2171 |
| | 12 | 100 | 95 | 168 | 174 | 115 | 51 | 15 | 1 | 0 | 0 | 0 | 719 |
| | 13 | 11 | 15 | 26 | 33 | 26 | 26 | 41 | 20 | 12 | 6 | 0 | 216 |
| | 14 | 0 | 3 | 3 | 4 | 5 | 0 | 3 | 3 | 7 | 8 | 5 | 41 |
| | Total | 2702 | 1163 | 608 | 353 | 170 | 81 | 60 | 24 | 19 | 14 | 5 | 5199 |

## 4 Summary and Discussion

Multiple wave hindcast products in the North Atlantic Ocean for 25 years are compared to the two wave buoy arrays distributed over the Northwest and Northeast Atlantic. The wave hindcast products used are TodaiWW3-NK, ERA5, IOWAGA(CFSR) and IOWAGA(ERA5). The sea state measured by the buoy array in the Northeast Atlantic is generally higher, but the spectral information is not available. For the comparison, significant wave height $H_s$ and mean wave period $T_{m02}$ are used because these two valuables represent the sea state. A special care is taken in the comparison of the mean wave period $T_{m02} = m_o/m_2$ because the higher spectral moment $m_2$ depends relatively more on the upper frequency limit, but the limit is not always the same among the buoy observations and wave hindcasts.

Significant wave heights $H_s$ from each of hindcast products are first compared to the wave buoy observations. All the four wave hindcast products show good agreement with the observations such that the correlation coefficients exceeded 0.9 for both cases for Northwest Atlantic and Northeast Atlantic. ERA5 showed slightly better performance than the other products, but the difference is limited in terms of the metrics such as bias, correlation coefficients, and RMSE. The difference between each model becomes larger and apparent when the high sea states ($H_s > 10$) are focused. The comparison of the probability density function of $H_s$ with the form of quantile-quantile plot, it was shown that the difference reaches almost 4m and 5m to the end in the Northwest Atlantic and Northeast Atlantic, respectively. For the extreme conditions, TodaiWW3-NK and IOWAGA(CFSR) showed better performance than the others. ERA5 tends to underestimate the extreme wave height, while IOWAGA(ERA5) tends to overestimate.

Prior to the systematic comparison of $T_{m02}$ between the hindcast products and the wave buoy observations, it was discussed how the integration range of the spectral moment $m_2$ influence on the mean wave period $T_{m02} = m_o/m_2$. The consideration of this point was motivated by the difference in the frequency range used by wave buoys as well as multiple hindcast products. For the quantitative description, observations in the Northwest Atlantic were examined in detail because some of the buoys consistently provide frequency spectra data. Changing the upper frequency limit of the integration from 1.0Hz to 0.4Hz turned out to change the mean of $T_{m02}$ by 10%. Some examples are shown that the double peak in the wave spectra leads to relatively large difference in $T_{m02}$ by the selection of the upper frequency limit. The influence of the upper frequency limit on the calculation of $T_{m02}$ becomes however smaller by increasing $H_s$ and the change becomes less than 3% if $H_s$ is larger than 10m. Following these results, the comparison of the mean wave period is limited to the high sea states ($H_s > 10$) because the difference in integration ranges less matters and because those wave conditions are more important for ship design. The conditional comparison of the mean wave period for the high sea states ($H_s > 10$) are conducted only for Northeast Atlantic because the number of samples is approximately 100 times less in the Northwest Atlantic. The results showed that all the hour wave hindcast underestimated the mean wave period.

The part of Scatter Diagram important for the extreme wave loads were referred as Scatter Diagram for Extreme conditions (SDE) and compared with the buoy observations. The range of SDE are $10 < H_s$ and $8 < T_{m02} < 14$ and so the comparison is again limited to the observations in Northeast Atlantic because such high sea state conditions are quite limited in the Northwest Atlantic. Compared to the SDE based on the buoy observations, SDEs based on four wave hindcast products showed large variation in the number of samples. More specifically, approximately 5000 cases were measured by the wave buoy arrays but TodaiWW3-NK reproduced approximately 7500 cases while ERA5 reproduced 2500 cases. IOWAGA(CFSR) and IOWAGA(ERA5) resulted in 6700 and 5200 cases, respectively. The large variations in the number of samples indicate that the wave hindcast requires some improvement to reproduce the extreme wave conditions. The maximum likelihood of $T_{m02}$ in each $H_s$ bin is also sometimes different between the buoy observation and the wave hindcast. Being consistent with the comparison of the mean wave period, the maximum likelihood of $T_{m02}$ is often found in the $T_{m02}$ bin one second longer in the observation than the wave hindcasts. The difference also indicates that the model uncertainty is relatively large for extreme conditions.

Based on the results described above, numerical wave model needs some improvement to reproduce the extreme conditions more accurately. The measurements of the extreme conditions are however in general limited because the probability of occurrence is low. The number of the samples needs to be increased to elucidate which part of the model needs further improvement and to conduct the validation. Using the satellite measurements should increase the data, but the wave period information is not obtained by the satellite altimeters. Recently developed distributed wave buoy sensor network [8] is potentially useful to measure the extreme conditions because of the wide coverage of measurements with the massive number of the sensors. Another type of the possible measurement method is the use of autonomous surface vehicle because the measurement position can be adjusted based on the forecast of the stormy events.

In the previous studies, the mean wave period $T_{m02}$ is often used just because $T_{m02}$ or equivalently $T_z$ is only available mean wave period measured by the wave buoys. However, $T_{m02}$ becomes sensitive to the upper limit of the frequency range as the wavelength becomes short as shown in this study. Since the upper limit has some variations among the wave buoys and numerical wave hindcasts, recalculation of $T_{m02}$ from the frequency spectra with the consistent frequency range is in theory necessary for comparison. Some considerations on the buoy response to ocean waves (i.e, RAO) are also necessary because the buoy response essentially works as low pass filtering. Perhaps using another mean wave period $T_{m01}$ instead of $T_{m02}$ is better because $T_{m01}$ is less sensitive to the high frequency range. Furthermore, the bulk wave statistics sometimes provide less meaning when the multiple wave systems such as the swell and wind coexist. In that case, the separation of the wave systems are necessary before calculating the bulk wave statistics [7]. Though the data volume is larger, the most accurate and direct comparison can be essentially made by the comparison of the frequency or directional wave spectra. More spectral data from the measurements needs to be shared especially in the area where the extreme wave conditions can occur.

## Acknowledgments

The study presented in this paper was carried out within a joint research project between the University of Tokyo and Nippon Kaiji Kyokai (ClassNK), and was funded by ClassNK. ERA5 data is available on the Copernicus Climate Change Service (C3S) Climate Data Store, and IOWAGA(CFSR) and IOWAGA(ERA5) data are retrieved from the IOWAGA data base (ftp://ftp.ifremer.fr/ifremer/ww3/HINDCAST//GLOBAL). Some wave spectra data are retrieved from NDBC (https://www.ndbc.noaa.gov/historical_data.shtml#stdmet).

## Declaration of compering interest

The authors declare that they have no known competing financial interests or personal relationships that could have appeared to influence by the report in this paper.